\documentclass[reprint,amsmath,amssymb,aip,jcp,floatfix,
               superscriptaddress]{revtex4-1}
\usepackage{times,color,graphicx,bm,multirow}
\usepackage[utf8]{inputenc}
\usepackage[colorlinks=true,urlcolor=blue,linkcolor=blue,
            citecolor=blue]{hyperref}
\usepackage{amsmath}

\newcommand{\bibtitle}[1]{\textit{#1},}

\mathchardef\mhyphen="2D

\begin{document}

\title{Optimizing Jastrow factors for the transcorrelated method}

\author{J. Philip Haupt}
  \affiliation{Max-Planck Institute for Solid State Research,
               Heisenbergstr.\ 1, 70569 Stuttgart, Germany}

\author{Seyed Mohammadreza Hosseini}
  \affiliation{Max-Planck Institute for Solid State Research,
               Heisenbergstr.\ 1, 70569 Stuttgart, Germany}

\author{Pablo L\'opez R\'ios}
  \email{p.lopez.rios@fkf.mpg.de}
  \affiliation{Max-Planck Institute for Solid State Research,
               Heisenbergstr.\ 1, 70569 Stuttgart, Germany}

\author{Werner Dobrautz}
  \affiliation{Department of Chemistry and Chemical Engineering,
               Chalmers University of Technology,
               41296 Gothenburg, Sweden}

\author{Aron Cohen}
  \affiliation{DeepMind, 6 Pancras Square, London N1C 4AG, UK}

\author{Ali Alavi}
  \email{a.alavi@fkf.mpg.de}
  \affiliation{Max-Planck Institute for Solid State Research,
               Heisenbergstr.\ 1, 70569 Stuttgart, Germany}
  \affiliation{Yusuf Hamied Department of Chemistry,
               University of Cambridge,
               Lensfield Road, Cambridge CB2 1EW, UK}

\begin{abstract}
We investigate the optimization of flexible tailored real-space
Jastrow factors for use in the transcorrelated (TC) method in
combination with highly accurate quantum chemistry methods such as
initiator full configuration interaction quantum Monte Carlo
(FCIQMC).
Jastrow factors obtained by minimizing the variance of the TC
reference energy are found to yield better, more consistent results
than those obtained by minimizing the variational energy.
We compute all-electron atomization energies for the challenging
first-row molecules C$_2$, CN, N$_2$, and O$_2$ and find that the TC
method yields chemically accurate results using only the cc-pVTZ basis
set, roughly matching the accuracy of non-TC calculations with the
much larger cc-pV5Z basis set.
We also investigate an approximation in which pure three-body
excitations are neglected from the TC-FCIQMC dynamics, saving storage
and computational cost, and show that it affects relative energies
negligibly.
Our results demonstrate that the combination of tailored real-space
Jastrow factors with the multi-configurational TC-FCIQMC method
provides a route to obtaining chemical accuracy using modest basis
sets, obviating the need for basis-set extrapolation and composite
techniques.
\end{abstract}

\maketitle

\section{Introduction}

The numerical study of the many-body electronic Schr{\"o}dinger
equation is plagued by a number of difficulties that originate from
the singular nature of the Coulomb potential, which at particle
coalescence points induces non-smooth behavior (cusps) in the wave
function. \cite{kato_cusp_1957, pack_cusp_1966}
This gives rise to slow convergence of quantum chemical methods with
respect to the size of the basis sets used.
One cure for this problem is to introduce explicit dependence in the
wave function on electron-electron and electron-nucleus distances.
\cite{hylleraas_explicit_1929}
This allows to analytically encode the non-smooth behavior into the
wave function ansatz, leaving only a relatively smooth wave function
for further treatment.
Within the so-called Jastrow ansatz, \cite{Jastrow_1955} the
electronic wave function is written as
\begin{equation}
  \label{eq:tc_wfn}
  \Psi = e^J \Phi \;,
\end{equation}
where $J = J({\bf r}_1,\ldots,{\bf r}_{N_{\rm e}})$ is a symmetric
correlation factor that contains optimizable parameters and depends on
the positions of the $N_{\rm e}$ electrons, while $\Phi$ is an
antisymmetric function, which can be taken to be a single Slater
determinant such as the Hartree-Fock (HF) state, or a
multi-configurational function such as a configuration-interaction
(CI) wave function.
With $J$ taking care of the majority of the non-analytic behavior,
$\Phi$ can be expected to be a relatively smooth function, accurately
expressible using small basis sets.

The presence of $J$ complicates the many-body integrals required to
solve the Schr\"odinger equation, which can usually only be handled by
the real-space variational and diffusion quantum Monte Carlo (VMC and
DMC) methods. \cite{Needs_casino_2020, Foulkes_QMC_2001}
The transcorrelated (TC) method of Boys and Handy
\cite{tc_boys_handy_1969, handy_be_1969, Handy_tcvar_1971} provides a
framework for the treatment of such an ansatz in second quantization
which requires only up to three-body integrals to be evaluated, but
involves a non-Hermitian effective Hamiltonian.
In the early work on the TC method, the non-Hermiticity of the
formalism was found to be highly problematic, and lead to a waning of
interest in the whole methodology. 

There was a revival of interest in the TC method in the late 1990s
with Nooijen and Bartlett, \cite{Nooijen_tc_1998} and Ten-no,
\cite{Tenno_mp2_2000, Tenno_cc_2000, Hino_2002} who also successfully
combined the TC method with quantum chemical methods such as
M{\o}ller-Plesset perturbation theory and the linearized coupled
cluster method.
Tsuneyuki, Umezawa, and co-workers developed a Slater-Jastrow
treatment of the TC Hamiltonian using the VMC method,
\cite{Umezawa_tcvmc_2003} and applied transcorrelation to solid-state
systems. \cite{Umezawa_tc_ueg_2004, Sakuma_tc_solids_2006,
Tsuneyuki_tc_solids_2008, Ochi_2013}

With the development of the R12/F12 class of explicitly correlated
methods, \cite{Klopper_r12_1987, Kutzelnigg_r12_1991,
Klopper_r12_1991, Noga_r12_1992, Noga_r12_1994, Klopper_r12_2002,
Valeev_r12_2004, Tenno_r12_2004, Kong_r12_review_2012} interest in the
TC method in quantum chemistry once again declined.
A second revival of the TC method has recently been spurred by the
realization that the non-Hermiticity of the TC Hamiltonian is
unproblematic \cite{Luo_tc_fciqmc_2018} for methods such as full
configuration-interaction quantum Monte Carlo (FCIQMC),
\cite{Guther_neci_2020, Booth_fciqmc_2009, Cleland_initiator_2010}
whilst providing substantial improvements in basis-set convergence
\cite{Cohen2019} and in the compactness of CI expansions, for example
for strongly-correlated two-dimensional Hubbard models.
\cite{Dobrautz_2D_Hubbard_2019}
Coupled-cluster methods have also been developed to treat the TC
Hamiltonian, both in its full form as well as via accurate
approximations for the efficient treatment of three-body interactions.
\cite{Liao2021, Schraivogel2021, Schraivogel2023, Christlmaier2023}
A consistent finding in these works has been that transcorrelation not
only improves basis-set convergence, as expected, but also the
effective level of theory of the underlying correlation method,
providing a major motivation to develop this methodology with the
ultimate aim of studying strongly-correlated ab initio systems. 

The TC method amounts to a non-unitary transformation of the
Hamiltonian of the system, and methods to solve such equations are in
general not guaranteed to converge to the exact total energy from
above as the complete basis set limit is approached.
The choice of Jastrow factor proves to be critical in the TC method,
since a poor choice can lead to highly non-variational TC energies and
poor error cancellation in energy differences.
In this paper we present a method for the optimization of Jastrow
factors catering to the TC approach, which we demonstrate in
combination with the FCIQMC method.

TC-FCIQMC is a stochastic eigensolver for the TC Hamiltonian, and
allows large CI solutions to be obtained for its eigenvectors.
This methodology has been used in the past to study first-row atoms
\cite{Cohen2019} using Boys-Handy-type Jastrow factors
\cite{boys_handy_form_1969} pre-optimized for use in the VMC method
\cite{Schmidt_atomic_Jastrows_1990}, the binding curve of the
beryllium dimer using a Boys-Handy Jastrow with an exponential kernel
optimized with VMC variance minimization \cite{Guther2021}, as well as
ultra-cold atomic systems with contact interactions.
\cite{Jeszenszki18, Jeszenszki20}
With the present method, new highly-flexible Jastrow factors tailored
to each system can be obtained for large molecules, allowing the
methodology to be extended, both within the context of TC-FCIQMC and
in other approaches such as TC density matrix renormalization group,
\cite{Liao_tc_dmrg_2023, Baiardi_tc_dmrg_2020} TC selected-CI
approaches, \cite{Ammar_tc_cipsi_2022} TC coupled-cluster theory,
\cite{Schraivogel2021, Liao2021} or for reducing the resources
required in a quantum computing setting.
\cite{Sokolov_tc_hubbard_2022, Dobrautz_tc_abinit_2022}
In our tests we find the atomization energies of various challenging
first-row molecules to be chemically accurate using the
moderately-sized standard cc-pVTZ basis set. \cite{Dunning_basis_1992}

The rest of this paper is structured as follows.
In Section \ref{sec:framework} we give an overview of the broader
theoretical framework used in our calculations.
Details of our proposed optimization methodology are given in Section
\ref{sec:method} along with data for various first-row atoms and
molecules to support our choices.
In Section \ref{sec:results} we analyze the accuracy of the TC method
in calculations of the atomization energies of these molecules which
we compare with their non-TC counterparts, and we present our
conclusions in Section \ref{sec:conclusions}.
Hartree atomic units ($\hbar=|e|=m_e=4\pi\epsilon_0=1$) are used
throughout unless stated otherwise.

\section{Methodological framework}
\label{sec:framework}

Substituting the Jastrow ansatz of Eq.\ \ref{eq:tc_wfn} into the
Schr\"odinger equation, $\hat H \Psi = E \Psi$, we obtain the
similarity-transformed Schr\"odinger equation,
\begin{equation}
  \label{eq:tc_schroedinger_eq}
  \hat{H}_{\rm TC} \Phi = E \Phi \;,
\end{equation}
where the transcorrelated Hamiltonian is
\begin{equation}
  \hat{H}_{\rm TC} = e^{-J} \hat{H} e^J \;,
\end{equation}
and $\Phi$ is the right-eigenvector of $\hat{H}_{\rm TC}$.
We approach solving the similarity-transformed Schr\"odinger equation
in two stages:\@ first we obtain a suitable $J$ using VMC-based
optimization, and then we obtain $\Phi$ as a CI wave function expanded
in a standard quantum chemical basis set using the TC-FCIQMC method.

The TC-FCIQMC method is based on the imaginary-time
Schr\"odinger equation with the TC Hamiltonian,
\begin{equation}
  -\frac{\partial \Phi}{\partial \tau} = \hat{H}_{\rm TC} \Phi \;,
\end{equation}
whose long-$\tau$ solution is the ground-state wave function that
satisfies Eq.\ \ref{eq:tc_schroedinger_eq}.
As in the non-TC FCIQMC algorithm, \cite{Booth_fciqmc_2009} $\Phi$ is
expressed as a general linear combination of Slater determinants,
\begin{equation}
  |\Phi\rangle = \sum_I c_I |D_I\rangle \;,
\end{equation}
where the $c_I$ coefficients are sampled by walkers in the FCIQMC
simulation.
Once the ground state has been reached, the total energy can be
evaluated by projection onto the HF determinant,
\begin{equation}
  \label{eq:E_proj}
  E_{\rm proj} =
    \frac {\langle D_{\rm HF}| {\hat H}_{\rm TC} |\Phi\rangle}
          {\langle D_{\rm HF} | \Phi\rangle} \;,
\end{equation}
which is averaged over time steps at finite walker numbers to obtain a
statistically meaningful result.

As has been shown by Luo and Alavi, \cite{Luo_tc_fciqmc_2018} the
non-Hermitian nature of $\hat{H}_{\rm TC}$ does not hamper the
convergence of the FCIQMC simulation, and the tools developed for that
method apply to $\hat{H}_{\rm TC}$ as they do to $\hat H$; this
includes the initiator approximation \cite{Cleland_initiator_2010}
which we use in all of the FCIQMC calculations reported in the present
work.
Furthermore, Dobrautz \textit{et al.\@}
\cite{Dobrautz_2D_Hubbard_2019} showed for the strongly-correlated 2D
Hubbard model that $\hat{H}_{\rm TC}$ has a significantly more compact
ground-state right eigenvector than $\hat H$ does, implying a
reduction in the initiator error, faster convergence with walker
number, and improved size consistency of the results.

The TC Hamiltonian can be exactly evaluated using the
Baker-Campbell-Hausdorff expansion, which for Jastrow factors dependent
only on electronic positions truncates exactly at second order,
\begin{equation}
  \hat{H}_{\rm TC} = e^{-J} \hat{H} e^J = \hat{H}
               + [\hat{H},J]
               + \frac{1}{2}[[\hat{H},J],J] \;.
\end{equation}
For a Jastrow factor containing up to two-electron contributions,
\begin{equation}
  J = \sum_{i<j} u({\bf r}_i, {\bf r}_j) \;,
\end{equation}
the explicit form of $\hat{H}_{\rm TC}$ is
\begin{eqnarray}
  \nonumber
  \hat{H}_{\rm TC}  & = & \hat{H}
    - \sum_{i}\left(\frac{1}{2}\nabla_{i}^{2}J
              + (\nabla_{i}J) \cdot \nabla_{i}
              + \frac{1}{2}(\nabla_{i}J)^{2}\right)\\
  & = & \hat{H} - \sum_{i<j}\hat{K}({\bf r}_{i},{\bf r}_{j})
       -\sum_{i<j<k} \hat{L}({\bf r}_{i},{\bf r}_{j},{\bf r}_{k}) \;,
\end{eqnarray}
where the additional two- and three-body terms are
\begin{eqnarray}
  \nonumber
  \hat K ({\bf r}_i,{\bf r}_j)
    & = & \frac 1 2 \left[
            \nabla_i^2 u({\bf r}_i,{\bf r}_j)
          + \nabla_j^2 u({\bf r}_i,{\bf r}_j) \right. \\ \nonumber
    & + &   \left.
            |\nabla_i u({\bf r}_i,{\bf r}_j)|^2
          + |\nabla_j u({\bf r}_j,{\bf r}_i)|^2 \right] \\ \nonumber
    & + &   \nabla_i u({\bf r}_i,{\bf r}_j) \cdot \nabla_i
          + \nabla_j u({\bf r}_i,{\bf r}_j) \cdot \nabla_j
  \;, \\ \nonumber
  \hat L ({\bf r}_i,{\bf r}_j,{\bf r}_k)
    & = & \nabla_i u({\bf r}_i,{\bf r}_j) \cdot
          \nabla_i u({\bf r}_i,{\bf r}_k) \\ \nonumber
    & + & \nabla_j u({\bf r}_j,{\bf r}_i) \cdot
          \nabla_j u({\bf r}_j,{\bf r}_k) \\
    & + & \nabla_k u({\bf r}_k,{\bf r}_i) \cdot
          \nabla_k u({\bf r}_k,{\bf r}_j) \;.
\end{eqnarray}
Using this first-quantized Hamiltonian, we can construct a
second-quantized Hamiltonian for a given set of orthonormal spatial
orbitals $\{\phi_1,\ldots,\phi_{n_{\rm orb}}\}$, with corresponding
fermionic spin-$\frac{1}{2}$ creation (annihilation) operators
$a^\dagger_{p\sigma}$ ($a_{p\sigma}$),
\begin{eqnarray}
  \nonumber
  \hat H_{\rm TC} & = &
    \sum_{pq\sigma} h^p_q a^\dagger_{p\sigma} a_{q\sigma} \\ \nonumber
    & + & \frac 1 2
            \sum_{pqrs} (V^{pq}_{rs}-K^{pq}_{rs})
            \sum_{\sigma \tau} a^\dagger_{p\sigma} a^\dagger_{q\tau}
                               a_{s\tau} a_{r\sigma} \\
    & - & \frac 1 6
            \sum_{pqrstu} L^{pqr}_{stu}
            \sum_{\sigma \tau \lambda}
            a^\dagger_{p\sigma} a^\dagger_{q\tau} a^\dagger_{r\lambda}
            a_{u\lambda} a_{t\tau} a_{s\sigma} \;,
\end{eqnarray}
where $h^p_q=-\frac 1 2 \langle\phi_p|\nabla^2|\phi_q\rangle$ and
$V^{pq}_{rs}=\langle\phi_p\phi_q|r_{12}^{-1}|\phi_r\phi_s\rangle$ are
the one- and two-body terms of the original Hamiltonian, and
$K^{pq}_{rs}=\langle\phi_p\phi_q|\hat{K}|\phi_r\phi_s\rangle$ and
$L^{pqr}_{stu}=\langle\phi_p\phi_q\phi_r|\hat{L}
|\phi_s\phi_t\phi_u\rangle$ are the corresponding terms arising from
the similarity transformation. \cite{Cohen2019}
Note that the 3-body operator $\hat{L}$ is Hermitian, and for real
orbitals the $L^{pqr}_{stu}$ tensor has 48-fold symmetry, a useful
property to reduce the storage requirement for these integrals.

By construction, Jastrow factors can be used to impose local Kato cusp
conditions \cite{kato_cusp_1957} based on the relative spin of
electron pairs, but are however ill-suited to describing more rigorous
electronic-state dependent cusp conditions.  \cite{pack_cusp_1966}
However, in the present work we use spin-independent Jastrow factors
for simplicitly, which we constrain to obey the opposite-spin
electron-electron cusp condition since it is physically more important
than the parallel-spin cusp condition.
The use of spin-dependent Jastrow factors would require replacing the
spatial-orbital indices above with spin-orbital indices, resulting in
an order of magnitude more three-body integrals to be computed and
stored.
Notwithstanding this increase in memory requirements, spin-dependent
Jastrow factors may offer other advantages such as faster basis set
convergence, and will be investigated in future work.

\section{Optimization methodology}
\label{sec:method}

In this section we present our methodological choices for the
optimization of Jastrow factors to be used in TC methods.
To illustrate the effect of these choices we compute the ground-state
energies of the all-electron C, N, and O atoms and the C$_2$, CN,
N$_2$, and O$_2$ molecules at their equilibrium geometries,
\cite{Feller_molecules_2008, Bytautas_diatomic_2005,
Harding_HEAT_2008} listed in Table \ref{table:bond_lengths}, with
non-TC and TC-FCIQMC calculations using HF orbitals (restricted
open-shell HF orbitals in the case of open-shell systems) expanded in
the standard cc-pV$x$Z family of basis sets \cite{Dunning_basis_1992}.
\begin{table}[htbp]
  \centering
    \caption{
    Electronic ground states and equilibrium bond lengths used for the
    molecules considered in this work, following Ref.\
    \onlinecite{Feller_molecules_2008}.
  }
  \label{table:bond_lengths}
  \begin{tabular}{ccc}
    System & State & $r_{\rm eq}$ ($\rm\AA$) \\
  \hline \hline
    C$_2$ & ${}^1\Sigma_g^+$ & $1.2425$ \\
    CN    & ${}^2\Sigma^+$   & $1.1718$ \\
    N$_2$ & ${}^1\Sigma_g^+$ & $1.0977$ \\
    O$_2$ & ${}^3\Sigma_g^-$ & $1.2075$ \\
  \hline
  \end{tabular}
\end{table}
The quality of energy differences is then assessed using the
atomization energy of the molecules.
In order to be able to determine whether the methodology is capable of
delivering chemically-accurate relative energies, \textit{i.e.},
incurring an error of less than $1$ kcal/mol $= 1.6$ mHa, we aim to
keep each of the different errors well below this threshold; we
comment on the magnitude of the expected error from each different
source in the subsections that follow.
We expect a total bias in our resulting relative energies of less
than $0.5$ mHa.

For all of our calculations we generate our orbitals and integration
grids using \textsc{pyscf}, \cite{pyscf} optimize Jastrow factors
using the \textsc{casino} continuum QMC package,
\cite{Needs_casino_2020} compute TC matrix elements using the
\textsc{tchint} library developed by the authors, \cite{tchint} and
perform FCIQMC and TC-FCIQMC calculations with the \textsc{neci}
FCIQMC package. \cite{Guther_neci_2020}
We report FCIQMC energies obtained by projection onto the HF
determinant, as per Eq.\ \ref{eq:E_proj}.

In the result of this section we will discuss the effect of modifying
each part of this calculation pipeline on the final TC-FCIQMC
energies; the aspects not being discussed are assumed to operate as
per our final recommendations unless otherwise stated, \textit{e.g.},
when we discuss the use of variance or energy minimization we use
integration grids of $l_{\rm grid}=2$, while when we discuss grid
density we use variance-minimized Jastrow factors.
In particular, note that the reported FCIQMC energies have been
extrapolated to the full CI (FCI) limit as described in Section
\ref{sec:tc_nw_conv} unless explicitly stated otherwise.

\subsection{Jastrow factor}

In continuum quantum Monte Carlo calculations the Jastrow factor for a
molecule consisting of $N_{\rm n}$ nuclei and $N_{\rm e}$ electrons is
usually constructed as the sum of isotropic electron-electron,
electron-nucleus, and electron-electron-nucleus terms,
\cite{LopezRios_Jastrow_2012}
\begin{equation}\label{eq:jastrow}
  \begin{split}
    J & = \sum_{i<j}^{N_{\rm e}} u(r_{ij})
        + \sum_i^{N_{\rm e}} \sum_I^{N_{\rm n}} \chi(r_{iI}) \\
      & + \sum_{i<j}^{N_{\rm e}} \sum_I^{N_{\rm n}}
            f(r_{ij},r_{iI},r_{jI}) \;.
  \end{split}
\end{equation}
As in Ref.\ \onlinecite{Drummond_Jastrow_2004} we express each of
these terms as a natural power expansion in the relevant
inter-particle distances,
\begin{equation}
  \begin{split}
    u(r_{ij})    & = t(r_{ij},L_u)
                     \sum_{k} a_k r_{ij}^k \;, \\
    \chi(r_{iI}) & = t(r_{iI},L_\chi)
                     \sum_{k} b_k r_{iI}^k \;, \\
    f(r_{ij}, r_{i}, r_{j}) & = t(r_{iI},L_f) t(r_{jI},L_f)
                     \sum_{k,l,m} c_{klm}
                        r_{ij}^k r_{iI}^l r_{jI}^m \;, \\
  \end{split}
\end{equation}
where $\{a_k\}$, $\{b_k\}$, and $\{c_{klm}\}$ are linear parameters,
$L_u$, $L_\chi$, and $L_f$ are cut-off lengths, $t(r,L) = (1-r/L)^3
\Theta(r-L)$ is a cut-off function, and $\Theta(r-L)$ is the Heaviside
step function.

In essence, the VMC and DMC methods sample real-space electronic
configurations $\{\bf R\}$ following an appropriate distribution based
on an analytic trial wave function $\Psi_{\rm T}({\bf R})$ and produce
a variational estimate of the total energy which is an average of the
local energy, $E_{\rm L} ({\bf R})= \Psi_{\rm T}^{-1}({\bf R}) \hat
H({\bf R}) \Psi_{\rm T}({\bf R})$, at the sampled configurations.
Both the electron-electron and electron-nucleus Kato cusp conditions
\cite{kato_cusp_1957} are of crucial importance in suppressing extreme
outliers in the local energy samples, making it possible to obtain
meaningful wave-function parameter sets from VMC-based optimization.
It is standard practice to apply the electron-electron cusp condition
on the $u$ term of the Jastrow factor, while the electron-nucleus cusp
condition is enforced by modifying the $l=0$ component, $\phi(r)$, of
cuspless molecular orbitals near nuclei so that they exhibit a cusp.
\cite{Ma_cusp_2005}
This has been found to be a better approach than applying the
electron-nucleus cusp condition via the parameters in the $\chi$ term
of the Jastrow factor. \cite{Drummond_Jastrow_2004, Ma_cusp_2005,
Needs_casino_2020}

However, in TC-FCIQMC it is preferable to use unmodified molecular
orbitals obtained from standard basis sets.
It would be possible to optimize the Jastrow factor parameters in VMC
in the presence of cusp-corrected orbitals and use them in TC-FCIQMC
with a cusp-uncorrected orbital basis, but the Jastrow factor would
then be sub-optimal by construction in the latter calculation.
Instead, we recast the cusp-correction scheme of Ref.\
\onlinecite{Ma_cusp_2005} as an electron-nucleus Jastrow factor term
$\Lambda$, to be added to (rather than replacing) the $\chi$ term in
Eq.\ \ref{eq:jastrow}.

We construct our cusp-correcting Jastrow factor term as
\begin{equation}\label{eq:cusp-corr-1}
  \begin{split}
    \Lambda(r) & = \left[ \ln \tilde \phi(r) - \ln \phi(r) \right]
                   \Theta(r-r_{\rm c}) \;,
  \end{split}
\end{equation}
where, using the notation of Ref.\ \onlinecite{Ma_cusp_2005}, $r_{\rm
c}$ is a cut-off radius, $\phi(r)$ is the $l=0$ component of the
target orbital at the desired nucleus, and $\tilde \phi(r)$ is its
cusp-corrected counterpart,
\begin{equation}\label{eq:cusp-corr-2}
  \tilde \phi(r) = e^{\sum_{l=0}^4 \lambda_l r^l} + C
                   \quad,\quad r<r_{\rm c} \;.
\end{equation}
Here, $\{\lambda_l\}$ are parameters determining the shape of the
corrected orbital and parameter $C$ is only set to a non-zero value in
the presence of nodes of $\phi(r)$ near the nucleus.

\begin{figure}[!hbt]
  \begin{center}
    \includegraphics[width=\columnwidth]{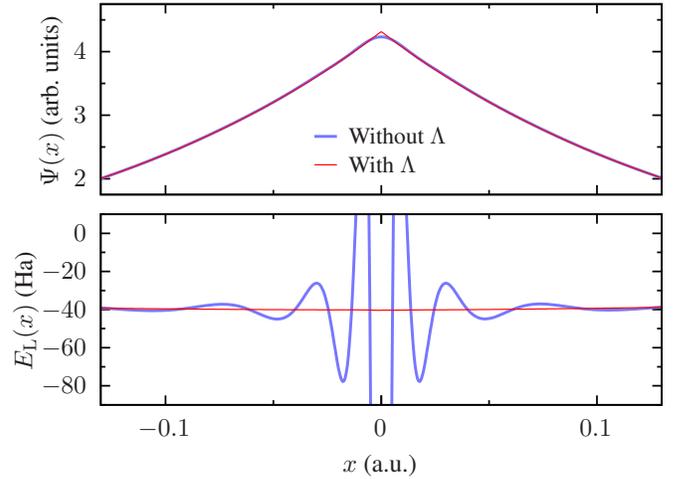}
    \caption{
      Wave function value and local energy as a function of the $x$
      coordinate of an electron in a carbon atom as it crosses the
      nucleus at $x=0$, both without and with a cusp-correcting
      $\Lambda$ Jastrow factor term applied to a HF wave function
      using the cc-pVDZ basis.
    }
    \label{fig:cusp-term}
  \end{center}
\end{figure}

Applying continuity and differentiability conditions (see Eq.\ 14 of
Ref.\ \onlinecite{Ma_cusp_2005}) leaves $\lambda_0$ and $r_{\rm c}$ as
the only free parameters in Eqs.\ \ref{eq:cusp-corr-1} and
\ref{eq:cusp-corr-2}.
Reference \onlinecite{Ma_cusp_2005} describes an approach to obtain
reasonable values for these parameters, which we use as initial values
to be refined by the subsequent VMC optimization procedure.
In practice we evaluate $\phi(r)$ in $\Lambda(r)$ by spline
interpolation of tabulated data.
Figure \ref{fig:cusp-term} illustrates the effect of using a $\Lambda$
term in practice.

For reference, in our calculations we optimize a total of $44$ Jastrow
factor parameters for the atoms and homonuclear dimers, and $80$
parameters for the CN dimer; we keep the $L_u$, $L_\chi$, and $L_f$
cut-off lengths fixed at sensible values for simplicity.

\subsection{Jastrow factor optimization}

The VMC method is not only capable of evaluating the variational
energy associated with a trial wave function $\Psi_{\rm T}$,
\begin{equation}
  E_{\rm VMC} = \frac { \langle \Psi_{\rm T}| \hat H
                        |\Psi_{\rm T} \rangle }
                      { \langle \Psi_{\rm T}|
                        \Psi_{\rm T} \rangle } \;,
\end{equation}
but also provides a variational framework in which any parameters $\bm
\alpha$ present in this trial wave function can be optimized; note
that in practice $|\Psi_{\rm T}\rangle = e^J |D_{\rm HF}\rangle$, and
$\bm \alpha$ are the optimizable parameters in $J$ in this work.
Wave function optimization is usually carried out using a
correlated-sampling approach in which a set of $n_{\rm opt}$
electronic real-space configurations ${\{\bf R}_{i}\}_{i=1}^{n_{\rm
opt}}$ distributed according to the initial wave function squared
$|\Psi_{\rm T}({\bf R}; {\bm \alpha}_0)|^2$ is generated, and then a
target function is minimized by varying $\bm \alpha$ at fixed ${\{\bf
R}_{i}\}$.
One such target function is the ``variance of the VMC energy,''
\cite{Umrigar_varmin_1988, Kent_varmin_1999}
\begin{equation}
  \sigma_{\rm VMC}^2 =
    \frac { \langle \Psi_{\rm T}| (\hat H - E_{\rm VMC})^2
            |\Psi_{\rm T} \rangle}
          {\langle \Psi_{\rm T}| \Psi_{\rm T} \rangle} \;,
\end{equation}
which reaches its lower bound of zero when the trial wave function is
an eigenstate of the Hamiltonian.
In practice, minimizing $\sigma_{\rm VMC}^2$ yields variational
energies affected by large random fluctuations, as demonstrated below.
In continuum QMC methods, modifications have been devised to
circumvent this problem, such as weight limiting, unreweighted
variance minimization, or the minimization of other measures of spread
such as the median absolute deviation from the median energy.
\cite{Needs_casino_2020}
The computational cost of optimizing Jastrow factors within VMC scales
as a small power of system size, typically estimated to be ${\cal
O}(N_{\rm e}^3)$.

\subsubsection{Minimizing the variance of the reference energy}

In the context of the TC method, the reference energy
\begin{equation}
  E_{\rm ref} = \langle D_{\rm HF}|
                  e^{-J} \hat H e^J
                |D_{\rm HF} \rangle \;,
\end{equation}
is of particular significance since it represents the starting point
of the calculation, \textit{e.g.\@}, it is the energy at $\tau=0$ of a
TC-FCIQMC calculation, or the zeroth-order contribution to the TC
coupled cluster energy.
We refer to its associated variance,
\begin{equation}
  \label{eq:var_eref_1}
  \sigma_{\rm ref}^2 =
    \langle D_{\rm HF}|
      e^{-J} |\hat H-E_{\rm ref}|^2 e^J
    |D_{\rm HF} \rangle \;,
\end{equation}
as the ``variance of the reference energy,'' which can be easily
evaluated for a finite VMC sample of size $n_{\rm opt}$ as the sample
variance of the Slater-Jastrow energy over the Hartree-Fock
distribution,
\begin{equation}
  S_{\rm ref}^2 =
    \frac 1 {n_{\rm opt}-1}
    \sum_{n=1}^{n_{\rm opt}}
      \left| \frac {\hat H({\bf R}_n) \Psi_{\rm SJ}({\bf R}_n)}
                   {\Psi_{\rm SJ}({\bf R}_n)} - {\bar E}_{\rm ref}
      \right|^2 \;,
\end{equation}
which tends to $\sigma_{\rm ref}^2$ as $n_{\rm opt}\to\infty$,
where $\Psi_{\rm SJ} = e^J D_{\rm HF}$ is the Slater-Jastrow wave
function, $\{{\bf R}_n\}_{n=1}^{n_{\rm opt}}$ are electronic
configurations distributed according to $D_{\rm HF}^2$, and the VMC
estimate of the reference energy is
\begin{equation}
  {\bar E}_{\rm ref} =
    \frac 1 {n_{\rm opt}}
    \sum_{n=1}^{n_{\rm opt}}
      \frac {\hat H({\bf R}_n) \Psi_{\rm SJ}({\bf R}_n)}
            {\Psi_{\rm SJ}({\bf R}_n)} \;.
\end{equation}
It should be noted that the variance of the reference energy has been
used as a target function for the optimization of Jastrow factors in
earlier work on the TC method, \textit{e.g.}, in Refs.\
\onlinecite{Handy_tcvar_1971} and \onlinecite{Umezawa_tcvmc_2003},
albeit in somewhat different theoretical frameworks.

To understand the physical significance of the variance of the
reference energy, note that Eq.\ \ref{eq:var_eref_1} can be written
in second quantized form as
\begin{equation}
  \sigma_{\rm ref}^2 =
    \sum_{I\ne {\rm HF}}
      \left|
        \langle D_I| \hat{H}_{\rm TC} |D_{\rm HF} \rangle
      \right|^2 \;,
\end{equation}
where $I$ runs over a complete basis set.
Minimizing $\sigma_{\rm ref}^2$ amounts to minimizing the coupling of
the HF determinant with the remainder of the space, which in the
TC-FCIQMC method reduces the spawning rate from the HF determinant to
its connected excited-state determinants, increasing the amplitude of
the HF determinant in the resulting $\Phi$.
In other words, if the Slater-Jastrow wave function were an exact
eigenstate of $\hat H$, a TC-FCIQMC simulation starting from the HF
determinant would immediately converge to the strictly
single-determinant solution.
Although this ideal scenario cannot be achieved in practice, it
nevertheless illustrates the benefits of obtaining a relatively
single-reference CI solution by minimizing the variance of the
reference energy.
We expect that this increased single-reference character will benefit
other approaches such as transcorrelated coupled-cluster methods as
well.

\begin{figure}[!hbt]
  \begin{center}
    \includegraphics[width=\columnwidth]{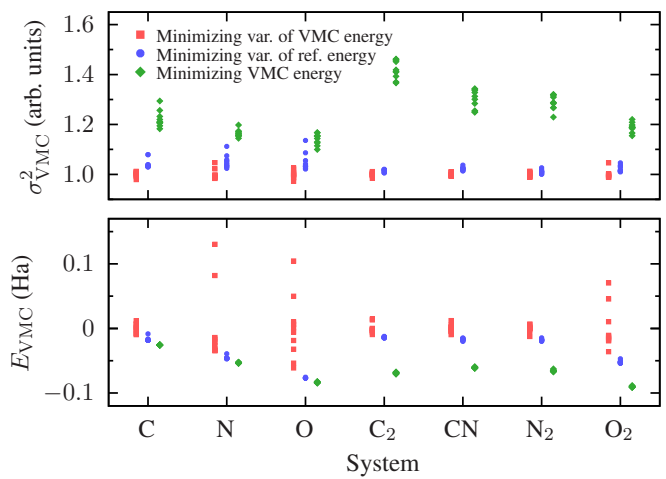}
    \caption{
      Variance of the VMC energy (top) and VMC energy (bottom) of
      the systems considered in this work using the cc-pVTZ basis and
      Jastrow factors obtained by minimizing the variance of the
      VMC energy (red squares), the variance of the reference energy
      (blue circles), or the VMC energy (green diamonds) in each of
      ten independent optimization runs with $n_{\rm opt}=10^5$ VMC
      configurations.
      To ease comparisons, variances have been rescaled and energies
      shifted by their average values from minimizing the variance
      of the VMC energy (\textit{i.e.}, the red squares average to a
      variance of $1$ and an energy of $0$ in the plot).
      The subpar ability of standard variance minimization to yield
      consistent VMC energies is evident in the bottom panel.
      \label{fig:varmin-E-Eref}}
  \end{center}
\end{figure}

We therefore investigate the performance of minimizing $\sigma_{\rm
ref}^2$ as an alternative to minimizing $\sigma_{\rm VMC}^2$ tailored
to the TC method.
In Fig.\ \ref{fig:varmin-E-Eref} we compare the VMC energy and
variance obtained by both variance minimization methods, along with
energy-minimized \cite{Nightingale_emin_2001, Toulouse_emin_2007,
Umrigar_emin_2007} results for reference, for the systems considered
in this work using $n_{\rm opt}=10^5$ VMC configurations.
Minimizing the variance of the VMC energy produces lower average
values of $\sigma_{\rm VMC}^2$, as one would expect, but also erratic
VMC energies with very large standard deviations (up to $\sim50$ mHa
in our tests).
Minimizing the variance of the reference energy, on the other hand,
produces values of $\sigma_{\rm VMC}^2$ which are only slightly higher
on average than those obtained from minimizing the variance of the VMC
energy ($1$--$5\%$ in our tests), while producing more stable VMC
energies with much smaller standard deviations (up to $\sim3$ mHa in
our tests).
We therefore do not use ``regular'' variance minimization since it
introduces large stochastic noise, making it unsuitable for optimizing
Jastrow factors, and from this point on we use the term ``variance
minimization'' to refer to the minimization of the variance of the
reference energy.

\subsubsection{Using an adequate sample size}

In continuum QMC calculations, Jastrow factors are usually optimized
using relatively few VMC configurations, with $n_{\rm opt}$ typically
in the tens of thousands.
However, it has been noted that substantially larger values of $n_{\rm
opt}$ in the hundreds of thousands are needed to converge expectation
values other than the VMC energy with respect to the wave function
parameters. \cite{Spink_trion_2016}

\begin{figure}[!hbt]
  \begin{center}
    \includegraphics[width=\columnwidth]{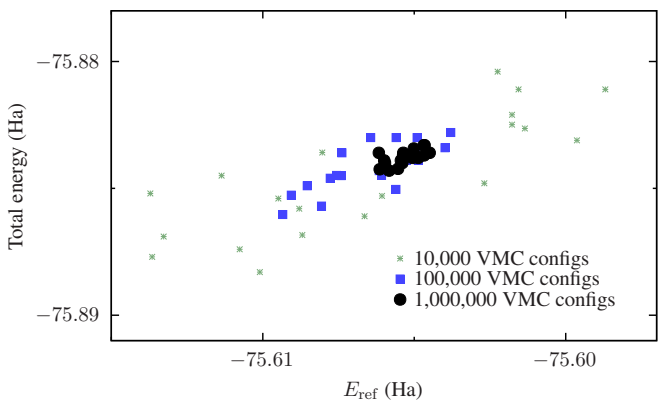}
    \caption{
      TC-FCIQMC energy of the C$_2$ molecule obtained using $n_{\rm
      w}=10^6$ walkers with the cc-pVDZ basis as a function of the reference
      energy $E_{\rm ref}$ for multiple independent Jastrow factor
      parameter sets obtained by variance minimization using three
      different VMC sample sizes.
      The horizontal spread is about $1.8$ times larger than the
      vertical spread, in line with the expectation that the standard
      deviation of the TC-FCIQMC energy is smaller than that of the
      reference energy.}
    \label{fig:spread_c2_cc-pvdz}
  \end{center}
\end{figure}

The level of convergence of any specific expectation value at a
specific value of $n_{\rm opt}$ can be easily assessed by performing
multiple optimization runs with different random seeds and evaluating
the standard deviation of the results.
In practice we find that we can use the uncertainty on the VMC
estimate of the reference energy $\bar E_{\rm ref}$ as a proxy for the
standard deviation on the TC-FCIQMC energy.
This is a reasonable replacement because
(i) the standard deviation of the TC-FCIQMC energy is not larger than
the standard deviation of the reference energy, as shown in Fig.\
\ref{fig:spread_c2_cc-pvdz} (it is usually significantly smaller
thanks to the ability of TC-FCIQMC to compensate for the presence of
a bias in $E_{\rm ref}$ via the correlation energy), and
(ii) the standard deviation of the reference energy is not larger than
the statistical uncertainty on the VMC estimate of the reference
energy $\bar E_{\rm ref}$ obtained with $n_{\rm opt}$ configurations
(it is usually significantly smaller due to the use of correlated
sampling in the optimization procedure).
We have verified that these inequalities hold for all the systems
studied in this work.

For the atoms and molecules considered in the present work we find
that $n_{\rm opt} = 2\times 10^7$ yields TC-FCIQMC energies with
standard deviations of less than $0.1$ mHa.
While this value of $n_{\rm opt}$ is three orders of magnitude greater
than typical values used in ``regular'' VMC calculations, the
optimization stage typically takes tens of core-hours, representing an
insignificant part of the total computational expense of TC-FCIQMC
runs.

\subsubsection{Energy minimization}

The obvious alternative to variance minimization is minimizing the VMC
energy itself, \cite{Nightingale_emin_2001, Toulouse_emin_2007,
Umrigar_emin_2007} which as demonstrated in Fig.\
\ref{fig:varmin-E-Eref} results in lower VMC energies but higher VMC
variances.
Energy minimization yields wave functions which minimize the
statistical fluctuations of the local energy in DMC calculations.
\cite{Ceperley_staterror_1986} and is typically the optimization
method of choice in the context of continuum QMC methods, but for
other purposes it is unclear whether the resulting wave functions
provide a better description of the system than those produced by
variance minimization.

\begin{figure}[!hbt]
  \begin{center}
    \includegraphics[width=\columnwidth]{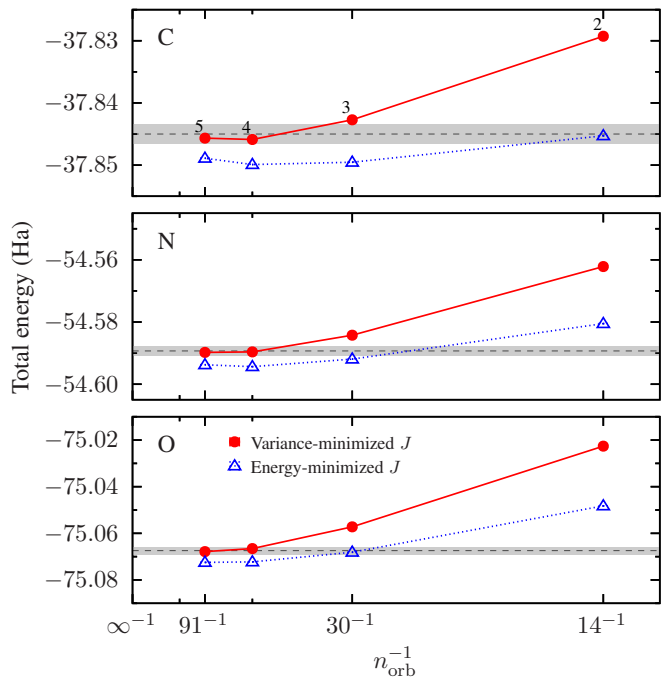}
    \caption{
      Total energy of the C, N, and O atoms as a function of the
      reciprocal of the number of molecular orbitals in the cc-pV$x$Z
      basis set.
      The non-variational behavior of up to about $5$ mHa is evident
      for the energy-minimized Jastrow factors, for which convergence to
      the exact energy as a function of basis-set size is rather slow.
      The shaded areas represent $\pm 1$~kcal/mol around the exact
      non-relativistic total energy from Ref.\
      \onlinecite{Bytautas_diatomic_2005}.
      Points in the top panel are annotated with the basis set
      cardinal number $x$.
    }
    \label{fig:bsdep-atoms-emin}
  \end{center}
\end{figure}

In Fig.\ \ref{fig:bsdep-atoms-emin} we compare the convergence with
basis-set size of TC-FCIQMC total energies of the C, N, and O atoms
using energy- and variance-minimized Jastrow factors.
Variance minimization appears to produce wave functions which converge
quickly and largely variationally to the basis set limit, while
energy-minimized wave functions tend to yield non-variational
TC-FCIQMC energies which converge very slowly to the basis set
limit.

\begin{figure}[!hbt]
  \begin{center}
    \includegraphics[width=\columnwidth]{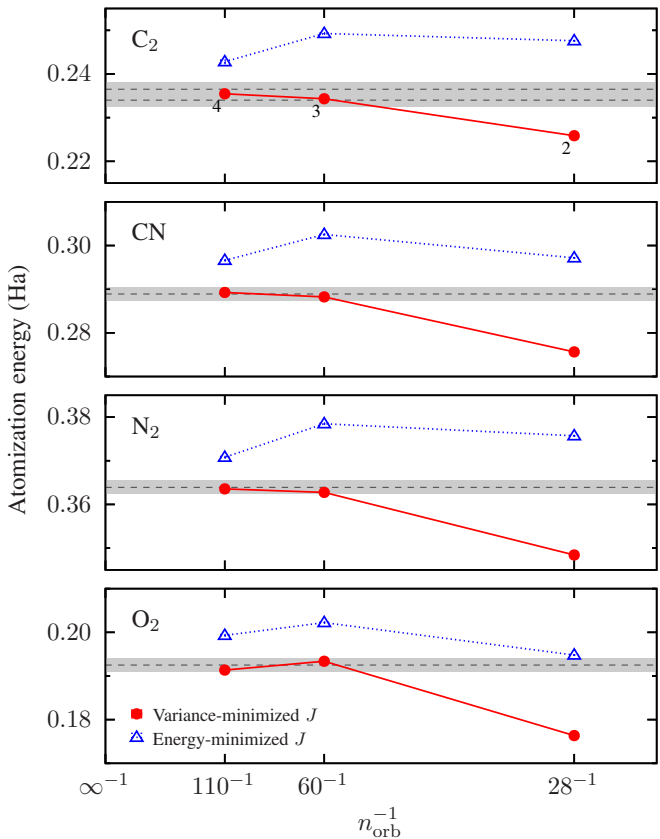}
    \caption{
      Atomization energy of the C$_2$, CN, N$_2$, and O$_2$ molecules
      as a function of the reciprocal of the number of molecular
      orbitals using the cc-pV$x$Z family of basis sets and Jastrow
      factors obtained by variance minimization and energy
      minimization.
      The FCI limit of the atomization energies for energy-minimized
      Jastrow factors are estimated from $10^6$-walker results
      assuming that the initiator error is identical to that using
      variance-minimized Jastrow factors at the same population.
      The shaded areas represent $\pm 1$~kcal/mol around the
      theoretical estimate of the non-relativistic atomization energy
      of Ref.\ \onlinecite{Feller_molecules_2008}; the distict
      estimate of Ref.\ \onlinecite{Bytautas_diatomic_2005} is also
      shown for C$_2$.
      Points in the top panel are annotated with the basis set
      cardinal number $x$.
    }
    \label{fig:bsdep-dimers-emin}
  \end{center}
\end{figure}

In Fig.\ \ref{fig:bsdep-dimers-emin} we plot atomization energies of
the C$_2$, CN, N$_2$, and O$_2$ molecules as a function of reciprocal
basis-set size, again demonstrating that variance-minimized Jastrow
factors exhibit favorable convergence properties.
Taking into account the evidence depicted in Figs.\
\ref{fig:bsdep-atoms-emin} and \ref{fig:bsdep-dimers-emin} we use
variance minimization to optimize our Jastrow factors for subsequent
use with the TC method.

\subsection{Matrix element evaluation}

Once the parameters in the Jastrow factor have been obtained, we
proceed to compute the TC contributions to the two- and three-body
terms of the Hamiltonian, $K_{ij}^{ab}$ and $L_{ijk}^{abc}$
respectively, by numerical integration on a grid.
For this task we use Treutler-Ahlrichs integration grids,
\cite{Becke_grids_1988, Treutler_grids_1995} which are atom-centred
grids constructed as the combination of a radial grid running up to
the Bragg radius and a Lebedev angular grid.
This type of integration grid is commonly used in density functional
theory calculations for which grids are usually ``pruned'' by reducing
the number of angular grid points near the atoms since typical
integrands are spherically symmetric around them, but this is not the
case for our integrands so we use ``unpruned'' grids.

We obtain the integration grids using \textsc{pyscf}, \cite{pyscf}
which provides an integer parameter, $l_{\rm grid}$, controlling
overall grid density.
We test for grid errors by evaluating TC-FCIQMC energies at $l_{\rm
grid}=0$--$5$ and defining the integration error as the difference of
each of these results with the value obtained by linear extrapolation
of the energies at $l_{\rm grid}=2$--$5$ to the $1/n_{\rm grid} \to 0$
limit.

\begin{figure}[!htb]
  \begin{center}
    \includegraphics[width=\columnwidth]{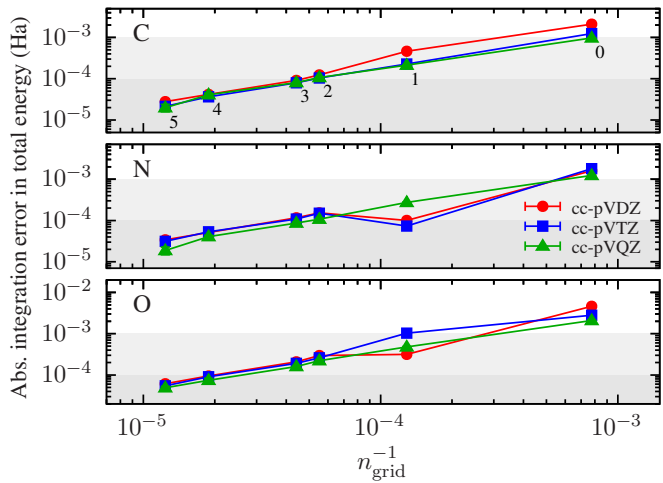}
    \caption{
      Absolute integration error in the total energy of the C, N, and
      O atoms as a function of $1/n_{\rm grid}$ using different basis
      sets in the cc-pV$x$Z family.
      The gray areas correspond to integration errors of less than
      $1$ and $0.1$ mHa.
      Points in the top panel are annotated with the value of
      \textsc{pyscf}'s grid density parameter $l_{\rm grid}$.
      Notice the logarithmic scale on both axes.}
    \label{fig:griderr-atoms}
  \end{center}
\end{figure}

First we investigate whether using basis sets of higher cardinal
number require finer grid densities to handle the sharper features in
the corresponding orbitals.
In Fig.\ \ref{fig:griderr-atoms} we plot the absolute integration
error in the total energy of the C, N, and O atoms as a function of
$1/n_{\rm grid}$ using the cc-pVDZ, cc-pVTZ, and cc-pVQZ basis sets.
If larger basis sets incurred a larger integration error one would
expect the cc-pVQZ results to consistently exhibit larger integration
errors in Fig.\ \ref{fig:griderr-atoms}, but instead we find that the
choice of basis set has little to no effect on the integration error,
indicating that the grid density need not be adjusted when the basis
set changes.

\begin{figure}[!htb]
  \begin{center}
    \includegraphics[width=\columnwidth]{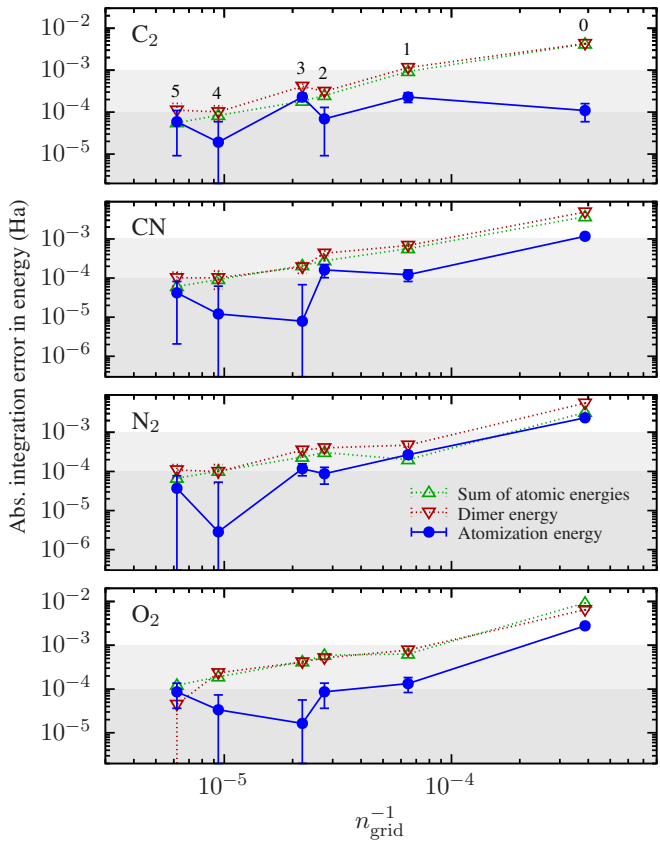}
    \caption{
      Absolute integration error in the energy as a function of the
      reciprocal of the number of grid points used for the evaluation
      of the TC integrals.
      Results are shown for the C$_2$, CN, N$_2$, and O$_2$ molecules
      at $10^5$ walkers using the cc-pVDZ basis set, along with the
      absolute integration error in the sum of the total energies of
      the atoms conforming each molecule at walker number convergence,
      and the absolute integration error in the atomization energy of
      the molecule obtained as the difference between both total
      energies.
      The gray areas correspond to integration errors of less than
      $1$ and $0.1$ mHa.
      Points in the top panel are annotated with the value of
      \textsc{pyscf}'s grid density parameter $l_{\rm grid}$.
      Notice the logarithmic scale on both axes.
      These results demonstrate that $l_{\rm grid}=2$ is sufficient to
      achieve sub-mHa accuracy in total energies and sub-$0.1$-mHa
      accuracy in relative energies.
    }
    \label{fig:griderr-dimers}
  \end{center}
\end{figure}

We now focus on the convergence of the integration error with grid
point density in total energies and in energy differences.
In Fig.\ \ref{fig:griderr-dimers} we plot the absolute integration
error in the total energies of the molecules, the atoms that conform
them, and in the corresponding atomization energies using the cc-pVDZ
basis.
We find substantial integration-error cancellation in energy
differences, with the atomization energy for all four molecules
reaching $0.1$ mHa for the $l_{\rm grid}=2$ grid, which is a
$60$-point radial grid combined with a $302$-point angular grid
totalling 18120 points per atom, which we therefore use throughout
this work.
Total energies incur integration errors of less than $1$ mHa for all
systems considered using $l_{\rm grid}=2$.

\subsection{TC-FCIQMC calculations}

\subsubsection{Convergence with the number of walkers}
\label{sec:tc_nw_conv}

The initiator approximation \cite{Cleland_initiator_2010} is a tool to
prevent uncontrolled walker growth in FCIQMC calculations, allowing to
stabilize the walker population around an arbitrary target value
$n_{\rm w}$.
This approximation incurs a bias with respect to the FCI limit which
goes to zero as $n_{\rm w}\to\infty$.
Typically the target population is increased until the FCIQMC energy
appears to converge to the target precision.
We find this approach to work remarkably well for the atoms considered
in this work, for which we use $n_{\rm w}=5\times10^6$ and
$5\times10^5$ in our non-TC and TC-FCIQMC calculations, respectively,
but for the molecules we find the level of convergence up to $n_{\rm
w} = 10^8$ to be unsatisfactory for our purposes.

For the semi-stochastic heat-bath CI (SHCI) method it has been shown
that one can obtain accurate estimates of total energies in the FCI
limit by linear extrapolation of SHCI energies to the limit in which
the second-order perturbation-theory (PT) correction $E_2$ is zero.
\cite{Smith_shci_extrap_2017, Holmes_shci_extrap_2017}
In order to obtain benchmark energies to compare FCIQMC results
against, we perfom SHCI calculations using the \textsc{dice} code
\cite{Sharma_dice1_2017, Holmes_dice2_2016} for the molecules
considered in this work with variational spaces of sizes such that
$|E_2|$ is not much larger than about $10$ mHa;
\cite{Smith_shci_extrap_2017, Holmes_shci_extrap_2017} we only exceed
this threshold for the cc-pV5Z calculations due to memory constraints.

\begin{figure}[!htb]
  \begin{center}
    \includegraphics[width=\columnwidth]{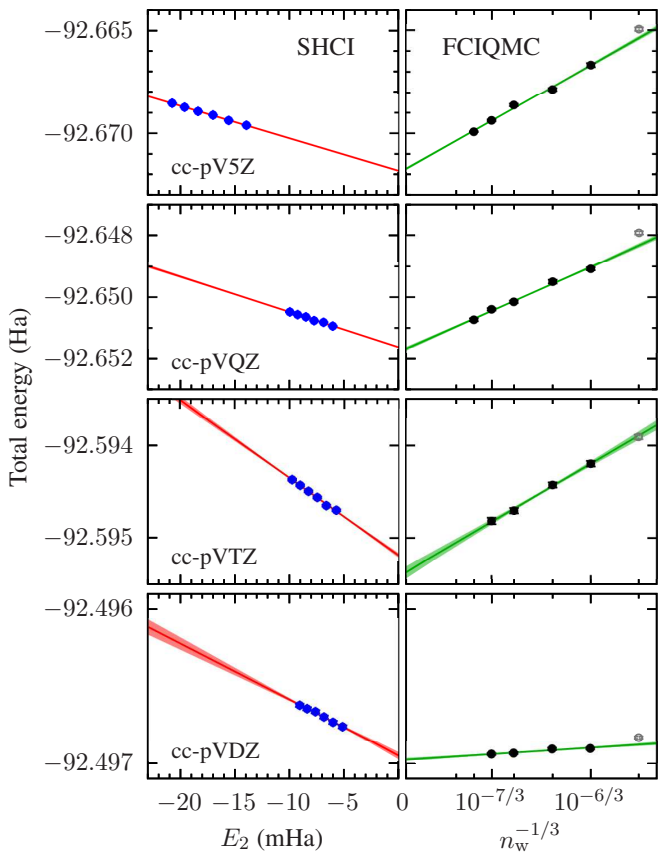}
    \caption{
      Extrapolation to the FCI limit of (non-TC) SHCI energies with
      respect to the second-order PT correction $E_2$ (left) and of
      FCIQMC energies with respect to $n_{\rm w}^{-1/3}$ (right) for
      the CN molecule.
      Grayed-out points are not used in the FCIQMC extrapolations.
      Translucent areas represent the uncertainty in the values of the
      fit at each point caused by the statistical uncertainty in the
      individual energies.
    }
    \label{fig:dual-extrap-cn}
  \end{center}
\end{figure}

We then perform non-TC FCIQMC calculations using between $10^6$ and
$2\times10^8$ walkers, and we empirically find that extrapolating the
FCIQMC energy linearly in $n_{\rm w}^{-1/3}$ to $n_{\rm w}^{-1/3}\to
0$ yields results in excellent agreement with the FCI limits from
SHCI.
We demonstrate this for the CN molecule in Fig.\
\ref{fig:dual-extrap-cn}; results for the other molecules are
similarly accurate, with FCI limits obtained from FCIQMC differing by
$0.4$ mHa or less from their SHCI counterparts throughout.
Note that in these extrapolations we do not include results at $n_{\rm
w} < 10^7$ walkers which we deem to lie ouside of the asymptotic
regime.

The theoretical foundation and validity of this extrapolation
technique will be studied in depth in future work.
Here we limit ourselves to empirically verifying that it yields
accurate FCI limits for the systems considered, and we make the sole
assumption that the technique continues to produce equally accurate
results for these molecules when the TC Hamiltonian is used.
We expect the extrapolation of TC-FCIQMC energies to require smaller
values of $n_{\rm w}$ than the non-TC case due to the increased
compactness of the CI wave function with the TC Hamiltonian.

The more compact the CI wave function is, the easier it is for FCIQMC
to sample the wave function accurately and the smaller the initiator
error becomes at fixed population.
The TC method has been shown to make CI wave functions more
compact for the two-dimensional Hubbard model.
\cite{Dobrautz_2D_Hubbard_2019}
Let $\{c_I\}$ be the $L^2$-normalized coefficients of the CI wave
function such that $\sum_I c_I^2=1$.
The quantity
\begin{equation}
  \label{eq:xi_compactness}
  \xi = \frac { c^{\rm (TC)}_{\rm HF} -
                c^{\rm (non\mhyphen TC)}_{\rm HF} }
              { 1 - c^{\rm (non\mhyphen TC)}_{\rm HF} } \;,
\end{equation}
is then a measure of the enhancement in the compactness of the wave
function, going from $0$ for no enhancement to $1$ if the TC wave
function becomes exactly single-determinantal.
From the data in Ref.\ \onlinecite{Dobrautz_2D_Hubbard_2019}, the TC
method yields a maximum $\xi = 0.64$ for the 18-site two-dimensional
Hubbard model.
We find that the values of $\xi$ for atomic and molecular systems are
not dissimilar from this; see Table \ref{table:l2_norms}.
\begin{table}[htbp]
  \centering
  \begin{tabular}{cccccccc}
          & C      & N      & O      &
            C$_2$  & CN     & N$_2$  & O$_2$  \\
  \hline \hline
  cc-pVDZ & $0.46$ & $0.63$ & $0.71$
          & $0.14$ & $0.23$ & $0.38$ & $0.53$ \\
  cc-pVTZ & $0.45$ & $0.61$ & $0.69$
          & $0.15$ & $0.24$ & $0.40$ & $0.55$ \\
  cc-pVQZ & $0.44$ & $0.60$ & $0.69$
          & $0.15$ & $0.24$ & $0.41$ & $0.57$ \\
  \hline
  \end{tabular}
  \caption{
    Enhancement of the compactness of the CI wave function, $\xi$ in
    Eq.\ \ref{eq:xi_compactness}, between our non-TC and TC FCIQMC
    calculations.
  }
  \label{table:l2_norms}
\end{table}

\begin{figure}[!htb]
  \begin{center}
    \includegraphics[width=\columnwidth]{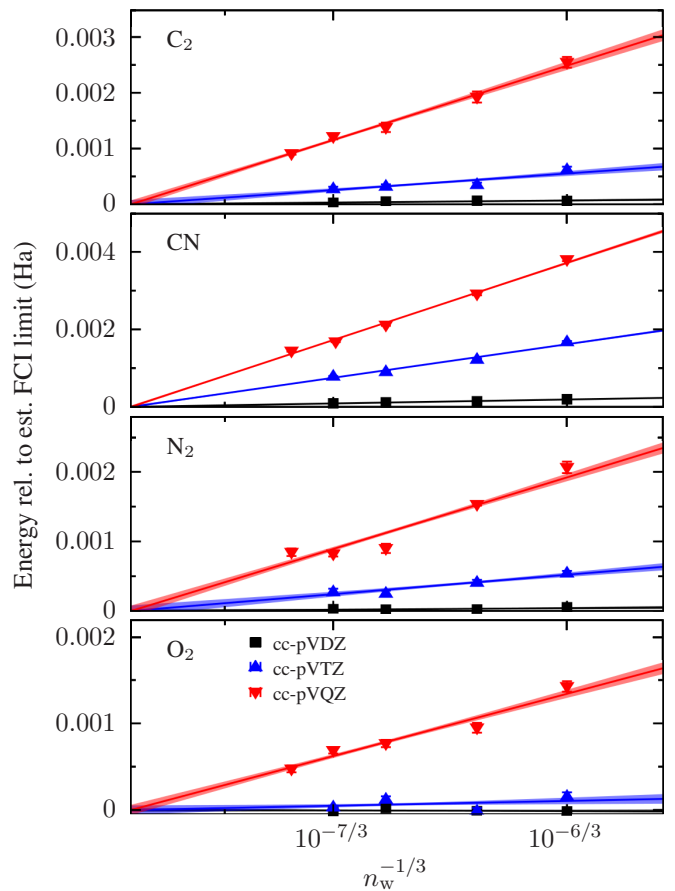}
    \caption{
      Estimated initiator error in the TC-FCIQMC energies as a
      function of $n_{\rm w}^{-1/3}$ for the C$_2$, CN, N$_2$, and
      O$_2$ molecules with the cc-pVDZ, cc-pVTZ, and cc-pVQZ basis
      sets.
      Translucent areas represent the uncertainty in the values of the
      fit at each point propagated from the statistical uncertainty in
      the individual energies.
    }
    \label{fig:extrap-dimers-tc}
  \end{center}
\end{figure}

Having established the more compact nature of the CI wave function
with the TC Hamiltonian, we deem appropriate to run the TC-FCIQMC
calculations using between $10^6$ and $10^7$ walkers for the cc-pVDZ
and cc-pVTZ bases, and between $10^6$ and $2\times 10^7$ walkers with
the cc-pVQZ basis.
In Fig.\ \ref{fig:extrap-dimers-tc} we plot the resulting TC-FCIQMC
energy relative to its corresponding FCI limit, \textit{i.e.}, the
initiator error, as a function of $n_{\rm w}^{-1/3}$ for each of the
molecules and basis sets considered.
As anticipated, the initiator error is to a good approximation
proportional to $n_{\rm w}^{-1/3}$, and we expect the extrapolated FCI
limits to incur errors no larger than $0.4$ mHa as in the non-TC case.

\subsubsection{Neglecting three-body excitations}

Sampling in FCIQMC calculations involves spawning walkers from
occupied determinants onto determinants connected to it by the
Hamiltonian.
The $\hat L$ matrix not only contributes to two-body excitations via
elements of the form $L_{ijk}^{ibc}$, but also connects determinants
which differ by a pure three-body excitation via $L_{ijk}^{abc}$,
which represents a huge increase in the connectivity of the Hilbert
space with respect to the non-TC problem.
However, the matrix elements $L_{ijk}^{abc}$ are usually very small in
magnitude, so neglecting pure three-body excitations is a valid
approximation to consider. \cite{Dobrautz_oneparam_2022}

This approximation effectively implies that $L_{ijk}^{abc}$ can be
regarded to be zero when all indices are distinct, except when three
or more indices correspond to spatial orbitals which are occupied in
the HF determinant, since these matrix elements are required during
the evaluation of the projected energy.
Therefore neglecting pure three-body excitations reduces the amount of
storage needed to hold $\hat L$ from ${\cal O}(n_{\rm orb}^6)$ to
${\cal O}(n_{\rm orb}^5) + {\cal O}(N_{\rm e}^3 n_{\rm orb}^3)$; we
report the specific reduction factors obtained for the molecules
considered in this work in Table \ref{table:no3_memory}.
\begin{table}[htbp]
  \centering
  \begin{tabular}{cccccccc}
          & C      & N      & O      &
            C$_2$  & CN     & N$_2$  & O$_2$  \\
  \hline \hline
  cc-pVDZ & $1.23$ & $1.17$ & $1.17$ &
            $1.87$ & $1.78$ & $1.78$ & $1.58$ \\
  cc-pVTZ & $2.04$ & $2.02$ & $1.93$ &
            $3.72$ & $3.66$ & $3.66$ & $3.46$ \\
  cc-pVQZ & $3.31$ & $3.44$ & $3.13$ &
            $6.60$ & $6.54$ & $6.57$ & $6.41$ \\
  \hline
  \end{tabular}
  \caption{
    $\hat L$ matrix storage reduction factor from neglecting
    pure three-body excitations, computed as the number of
    non-zero matrix elements in a the full $\hat L$ matrix
    divided by the number of non-zero matrix elements with
    repeated indices or three or more indices corresponding to
    orbitals occupied in the HF determinant.
  }
  \label{table:no3_memory}
\end{table}

Two-body excitations are in practice more expensive to attempt than
triple excitations, so neglecting the latter actually
\textit{increases} the cost per step of the calculation.
However, neglecting pure three-body excitations allows the TC-FCIQMC
time step to be larger, resulting in reduced serial correlation in the
statistics which enables reaching the target accuracy in fewer steps,
and one can generally expect a net cost reduction from this
approximation.
We report the specific reduction factors found for the molecules
considered in this work in Table \ref{table:no3_cost}.
\begin{table}[htbp]
  \centering
  \begin{tabular}{cccccccc}
          & C     & N     & O     &
            C$_2$ & CN    & N$_2$ & O$_2$ \\
  \hline \hline
  cc-pVDZ & $0.9$ & $1.0$ & $1.1$ &
            $1.7$ & $1.2$ & $1.5$ & $1.6$ \\
  cc-pVTZ & $1.0$ & $1.0$ & $0.8$ &
            $2.4$ & $1.0$ & $1.8$ & $2.0$ \\
  cc-pVQZ & $1.5$ & $1.5$ & $1.0$ &
            $3.1$ & $0.9$ & $1.9$ & $2.3$ \\
  \hline
  \end{tabular}
  \caption{
    Reduction factor in the walltime required to advance one unit of
    imaginary time at fixed population from neglecting pure three-body
    excitations in the TC-FCIQMC calculation.
    \label{table:no3_cost}}
\end{table}

In Table \ref{table:no3_Eat_error} we show the error in the
atomization energy of the molecules incurred by neglecting pure
three-body excitations.
\begin{table}[htbp]
  \centering
  \begin{tabular}{cllll}
                            &
  \multicolumn{1}{c}{C$_2$} &
  \multicolumn{1}{c}{CN   } &
  \multicolumn{1}{c}{N$_2$} &
  \multicolumn{1}{c}{O$_2$} \\
  \hline \hline
  cc-pVDZ   & $-0.62(2)$ & $-0.46(0)$ & $-0.56(2)$ & $-0.55(2)$ \\
  cc-pVTZ   & $-0.36(5)$ & $-0.30(2)$ & $-0.32(5)$ & $-0.20(3)$ \\
  cc-pVQZ   & $-0.45(6)$ & $-0.21(2)$ & $-0.32(7)$ & $-0.27(5)$ \\
  \hline
  \end{tabular}
  \caption{
    Error in the atomization energy of the molecules considered in this
    work incurred by neglecting pure three-body excitations from the
    FCIQMC dynamics, in mHa.
    \label{table:no3_Eat_error}}
\end{table}
We find that this approximation results in errors of the order of
$\sim0.3$ mHa at the cc-pVTZ level, which is a relatively small bias
considering the substantial storage and cost benefits of the
approximation.
Note that we do not use this approximation in the main results
presented in the next section.

\section{Results and discussion}
\label{sec:results}

We now analyze the accuracy of the TC method with tailored Jastrow
factors by comparing the convergence of the results as a function of
basis set size with non-TC results and with benchmark complete
basis-set limit (CBS) values from Refs.\
\onlinecite{Feller_molecules_2008, Bytautas_diatomic_2005,
Harding_HEAT_2008}.

In Table \ref{table:total_energies} we list the total energies that
we obtain for each system and basis set.
\begin{table*}[htbp!]
  \centering
  \begin{tabular}{c@{\;\;}|@{\;\;}crrrrrrr}
    \multicolumn{2}{c}{~}     &
    \multicolumn{1}{c}{C}     &
    \multicolumn{1}{c}{N}     &
    \multicolumn{1}{c}{O}     &
    \multicolumn{1}{c}{C$_2$} &
    \multicolumn{1}{c}{CN}    &
    \multicolumn{1}{c}{N$_2$} &
    \multicolumn{1}{c}{O$_2$} \\
    \hline \hline
    \multicolumn{9}{c}{~} \\[-0.4cm]
    \parbox[t]{2mm}{\multirow{5}{*}{\rotatebox[origin=c]{90}{non-TC}}}
    &cc-pVDZ& $ -37.7619$ & $ -54.4801$ & $ -74.9117$
            & $ -75.7320$ & $ -92.4970$ & $-109.2809$ & $-149.9915$ \\
    &cc-pVTZ& $ -37.7900$ & $ -54.5252$ & $ -74.9853$
            & $ -75.8094$ & $ -92.5954$ & $-109.4014$ & $-150.1554$ \\
    &cc-pVQZ& $ -37.8126$ & $ -54.5535$ & $ -75.0236$
            & $ -75.8578$ & $ -92.6517$ & $-109.4653$ & $-150.2362$ \\
    &cc-pV5Z& $ -37.8199$ & $ -54.5627$ & $ -75.0369$
            & $ -75.8752$ & $ -92.6717$ & $-109.4881$ & $-150.2655$ \\
    &cc-pV6Z& $ -37.8263$ & $ -54.5697$ & $ -75.0447$
            &             &             &             &             \\
    \multicolumn{9}{c}{~} \\[-0.3cm]
    \parbox[t]{2mm}{\multirow{4}{*}{\rotatebox[origin=c]{90}{TC}}}
    &cc-pVDZ& $ -37.8293$ & $ -54.5622$ & $ -75.0226$
            & $ -75.8844$ & $ -92.6671$ & $-109.4727$ & $-150.2216$ \\
    &cc-pVTZ& $ -37.8427$ & $ -54.5842$ & $ -75.0572$
            & $ -75.9197$ & $ -92.7152$ & $-109.5312$ & $-150.3078$ \\
    &cc-pVQZ& $ -37.8459$ & $ -54.5896$ & $ -75.0665$
            & $ -75.9272$ & $ -92.7247$ & $-109.5428$ & $-150.3244$ \\
    &cc-pV5Z& $ -37.8457$ & $ -54.5898$ & $ -75.0678$
            &             &             &             &             \\
    \multicolumn{9}{c}{~} \\[-0.3cm]
    \multicolumn{2}{c}{Ref.\ \onlinecite{Feller_molecules_2008}}
            &             &             &
            & $ -75.9240$ & $ -92.7232$ & $-109.5425$ & $-150.3273$ \\
    \multicolumn{2}{c}{Ref.\ \onlinecite{Bytautas_diatomic_2005}}
            & $ -37.8450$ & $ -54.5893$ & $ -75.0674 $
            & $ -75.9265$ &             & $-109.5427$ & $-150.3274$ \\
    \multicolumn{2}{c}{Ref.\ \onlinecite{Harding_HEAT_2008}}
            & $ -37.8450$ & $ -54.5893$ & $ -75.0674 $
            &             & $ -92.7229$ & $-109.5425$ & $-150.3275$ \\
    \hline
  \end{tabular}
  \caption{Total energies in Ha obtained for the atoms and molecules
    considered in this work, along with benchmark non-relativistic
    results.
    Statistical uncertainties arising from Monte Carlo sampling
    are smaller than $0.0001$ Ha in all cases.
  }
  \label{table:total_energies}
\end{table*}
We find the TC total energies to be remarkably accurate already at the
cc-pVQZ basis set level, differing by less than $2$ mHa per atom from
benchmark CBS values, while the non-TC total energies still miss the
benchmarks by $25$--$30$ mHa per atom with the cc-pV5Z basis set, and
$20$ mHa per atom at the cc-pV6Z level.
The TC total energies exhibit slight non-variational convergence,
with the atomic energies reaching values $0.5$ mHa below the benchmark
before increasing again towards it for larger basis-set sizes.
While non-variationality is not a desirable feature in a method, the
amount by which the TC results dip below the basis-set limit is
sufficiently tiny for this issue to be entirely ignored in practice.

\begin{figure}[!hbt]
  \begin{center}
    \includegraphics[width=\columnwidth]{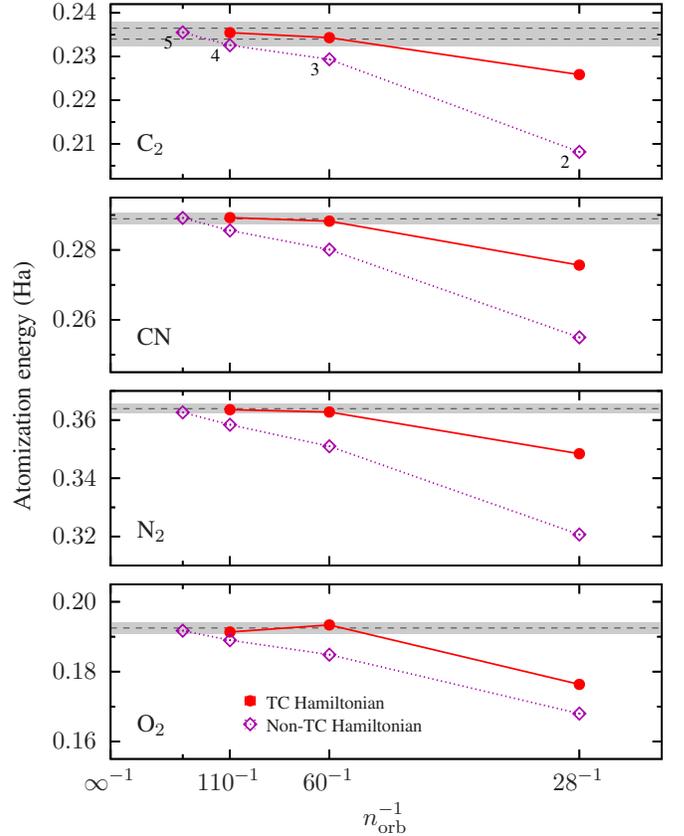}
    \caption{
      Atomization energy of the C$_2$, CN, N$_2$, and O$_2$ molecules
      obtained with FCIQMC and TC-FCIQMC as a function of the
      reciprocal of the number of molecular orbitals using the
      cc-pV$x$Z family of basis sets.
      Points in the top panel are annotated with the basis set
      cardinal number $x$.
      The shaded areas represent $\pm 1$~kcal/mol around the
      theoretical estimate of the non-relativistic atomization energy
      of Ref.\ \onlinecite{Feller_molecules_2008}; the distinct
      estimate of Ref.\ \onlinecite{Bytautas_diatomic_2005} is also
      shown for C$_2$.
    }
    \label{fig:bsdep-dimers-nontc}
  \end{center}
\end{figure}

From a chemical perspective, relative energies are more important than
total energies.
The atomization energies of the C$_2$, CN, N$_2$, and O$_2$ molecules
obtained from the total energies in Table \ref{table:total_energies}
are given in Table \ref{table:atomization} and plotted in Fig.\
\ref{fig:bsdep-dimers-nontc}.
\begin{table}[htbp!]
  \centering
  \begin{tabular}{c@{\;\;}|@{\;\;}crrrr}
    \multicolumn{2}{c}{}      &
    \multicolumn{1}{c}{C$_2$} &
    \multicolumn{1}{c}{CN}    &
    \multicolumn{1}{c}{N$_2$} &
    \multicolumn{1}{c}{O$_2$} \\
    \hline \hline
    \multicolumn{6}{c}{} \\[-0.4cm]
    \parbox[t]{2mm}{\multirow{4}{*}{\rotatebox[origin=c]{90}{non-TC}}}
    &cc-pVDZ& $208.2$ & $255.0$ & $320.7$ & $168.0$ \\
    &cc-pVTZ& $229.4$ & $280.1$ & $351.0$ & $184.9$ \\
    &cc-pVQZ& $232.6$ & $285.6$ & $358.4$ & $189.0$ \\
    &cc-pV5Z& $235.5$ & $289.2$ & $362.6$ & $191.7$ \\
    \multicolumn{6}{c}{} \\[-0.3cm]
    \parbox[t]{2mm}{\multirow{3}{*}{\rotatebox[origin=c]{90}{TC}}}
    &cc-pVDZ& $225.9$ & $275.6$ & $348.4$ & $176.4$ \\
    &cc-pVTZ& $234.3$ & $288.2$ & $362.7$ & $193.4$ \\
    &cc-pVQZ& $235.5$ & $289.3$ & $363.6$ & $191.4$ \\
    \multicolumn{6}{c}{} \\[-0.3cm]
    \multicolumn{2}{c}{Ref.\ \onlinecite{Feller_molecules_2008}}
            & $234.0$ & $288.9$ & $363.9$ & $192.5$ \\
    \multicolumn{2}{c}{Ref.\ \onlinecite{Bytautas_diatomic_2005}}
            & $236.5$ &         & $364.1$ & $192.6$ \\
    \multicolumn{2}{c}{Ref.\ \onlinecite{Harding_HEAT_2008}}
            &         & $288.6$ & $363.9$ & $192.7$ \\
    \hline
  \end{tabular}
  \caption{Atomization energies in mHa obtained for the molecules
    considered in this work, along with benchmark non-relativistic
    results.
    Statistical uncertainties arising from Monte Carlo sampling
    are smaller than $0.1$ mHa in all cases.
  }
  \label{table:atomization}
\end{table}
Again, we find that the TC results exhibit much faster convergence
with basis-set size than their non-TC counterparts, with TC
atomization energies being chemically accurate with respect to the CBS
benchmarks already at the cc-pVTZ basis-set level, matching the
accuracy of non-TC atomization energies obtained using the cc-pV5Z
basis set.
The TC method therefore provides an advantage of about two cardinal
numbers for the computation of relative energies.

The application of the TC method to quantum chemical methods in
general could be presumed to be problematic because any theoretical
guarantee of cancellation of errors in energy differences disappears
with the introduction of separately-optimized Jastrow factors for each
system.
However, the fact that in our results the relative energy converges at
smaller basis-set sizes than the total energy implies that substantial
error cancellation is at play in practice.
We take this as evidence that the TC method with tailored Jastrow
factors does not fully suppress advantageous error cancellation from
the underlying methodology.

It is important to note that neural-network based trial wave functions
proposed in recent years for use with VMC and DMC,
\cite{Pfau_ferminet_2020, Hermann_paulinet_2020} while promising, do
not achieve chemical accuracy reliably.
For example, an atomization energy of N$_2$ of $361.2(2)$ mHa is
obtained from the results provided by Ref.\
\onlinecite{Pfau_ferminet_2020}, which is $2.7(2)$ mHa away from the
benchmark; the slight difference in the bond length used for that
calculation only accounts for $0.1$ mHa of the energy difference.
By constrast, our TC-FCIQMC atomization energy obtained with the
cc-pVTZ basis is only $1.2$ mHa away from the benchmark.

\section{Conclusions}
\label{sec:conclusions}

We present a method to optimize flexible Jastrow factors of a form
commonly used in continuum quantum Monte Carlo methods for application
in the TC method, which we have tested within TC-FCIQMC.
Minimizing the variance of the reference energy is shown to be an
especially good fit for the TC method since it maximizes the
single-reference character of the CI wave function, and we have
demonstrated how this method outperforms standard energy minimization
in this context.
Various approximate aspects of the calculations have been
considered, and care has been taken to ensure that we can produce
relative energies within significantly less than $1$ mHa of their
FCI limit.

Our results show that the TC method with tailored Jastrow factors
delivers remarkably accurate total energies, and gives relative
energies with a cc-pVTZ basis set which rival the accuracy of non-TC
relative energies with the much larger cc-pV5Z basis set.
We expect that future work in this topic, which will likely include
technical enhancements, efficient approximations for dealing with the
three-body integrals, deterministic optimization methods, the use of
polyatomic and spin-dependent Jastrow factor terms, and methodological
adaptations for strongly correlated systems, will enable further
reductions in the basis-set sizes required to perform accurate quantum
chemistry calculations with the TC method.

\begin{acknowledgments}
The authors would like to thank H. Luo, D. Kats, K. Liao, and E.
Christlmaier for useful discussions, and K. Guther and K. Ghanem for
their early work on the \textsc{tchint} library.
P.L.R. and A.A. acknowledge support from the European Centre of
Excellence in Exascale Computing TREX, funded by the Horizon 2020
program of the European Union under grant no.\ 952165.
W.D. acknowledges funding from the Horizon Europe research and
innovation program of the European Union under the Marie
Sk{\l}odowska-Curie grant agreement no.\ 101062864.
Any views and opinions expressed are those of the authors only and do
not necessarily reflect those of the European Union or the European
Research Executive Agency.
Neither the European Union nor the granting authority can be held
responsible for them.
\end{acknowledgments}

\section*{Author declarations}
The authors have no conflicts to disclose.

\section*{Data availability}
The data that support the findings of this study are available from
the corresponding author upon reasonable request.



\begin{thebibliography}{10}
\bibitem{tchint}
  Transcorrelated Hamiltonian integral library \textsc{tchint} to be
  released; available from the authors upon reasonable request.
\end{thebibliography}
\end{document}